\documentclass[prl,aps,showpacs,twocolumn]{revtex4}
\usepackage{graphicx}
\begin{document}

\title{Superdecoherence through gate control noise}
\author{J\"urgen T. Stockburger}
\email{J.T.Stockburger@physik.uni-freiburg.de}
\altaffiliation[Current address: ]{Physikalisches Institut,
  Universit\"at Freiburg, Hermann-Herder-Str. 3, 79104 Freiburg, Germany}
\affiliation{Institut f\"ur Theoretische Physik II -
Universit\"at Stuttgart,
Pfaffenwaldring 57,
70550 Stuttgart, Germany}

\begin{abstract}
The external control circuits of quantum gates inevitably introduce a
small but finite noise to the operation of quantum computers. The
complex modes of decoherence introduced by this noise are not covered
by the common error models. Using the controlled-phase gate as an
example, the effect of gate control noise on decoherence is
investigated for different quantum computer architectures. It is shown
that the decoherence rate rises faster than linearly with the length
of a quantum register for most cases considered, adding to the
challenge of implementing proposed error correcting and fault tolerant
computation schemes. Sometimes an unwanted effective inter-qubit
coupling associated with the noise appears.
\end{abstract}


\maketitle

The concept of quantum computation, based on seminal ideas of Feynman
\cite{feynm85} has captivated the attention of a major part of the
physics community. Work in the field virtually exploded after it was
pointed out that an ideal quantum computer can solve certain problems
of practical interest faster than any classical computer by
an exponential factor.  Special features of quantum mechanics once
considered arcane, such as entanglement and non-locality, are now the
basis of proposed technologies.

Since quantum computers are essentially analog devices, they have to
be equipped with countermeasures against the effects of production
tolerances and of decoherence. The latter was recognized as the
crucial performance-limiting effect in quantum computers
\cite{unruh95}. The control of two-qubit quantum gates by external
(noisy) signals introduces a so far unexplored universal decoherence
mechanism whose rate scales less favorably with system size than in
cases previously considered.

Decoherence is the decay of interference which happens when a
deterministic phase relation is replaced by a random one
\cite{zurek91}.  Usually a {\em pointer variable} defines a preferred
basis in which the density matrix of a quantum system becomes diagonal
when decoherence is complete. The pointer variable also introduces an
important concept of `distance' of two quantum states in
superposition. In typical cases, the decoherence rate depends
quadratically on this distance.

In the context of quantum computation, decoherence is often analyzed
in terms of {\em error models}. In the simplest model, it is assumed
that different qubits are coupled to different environments, leading
to independent errors in different qubits. The decoherence rate then
scales at most linearly with the number $L$ of qubits in a quantum
register \cite{unruh95}, leaving the {\rm error rate} per qubit
constant when $L$ is increased. Often a fixed error rate per gate is
assumed as well. Based on this model, elaborate concepts for quantum
error correction \cite{stean96,knill97a,knill00} and fault-tolerant
computation \cite{shor96,knill98,knill05,svore05,aharo06} have been
pointed out. More recently, some of these ideas have been extended to
correlated errors \cite{kless06,novai06}.

For decoherence mechanisms in the other extreme, coupling of all
qubits in identical ways to a single environment, some quantum
superpositions decay with a rate proportional to $L^2$
(superdecoherence), while others are part of decoherence-free
subspaces allowed by the inherent symmetry of this kind of
qubit-environment coupling. A realistic quantum computation of any
appreciable size must then be performed using logical qubits residing
in decoherence-free subspaces \cite{zanar97,lidar98}. Similarly as in
quantum error correcting codes, several physical qubits are needed to
encode one logical qubit.

So far the decoherence mechanisms studied in the context of quantum
computation have been mainly due to noise sources inherent in the
technologies or materials used to implement single qubits or quantum
gates. Here we propose
to study a new decoherence mechanism, which emerges when a complete
architecture is analyzed rather than single components. The basic idea
is quite simple: To control the execution of a quantum algorithm,
quantum gates must be controlled by signals from external
circuits. The connection to the control circuit will be assumed
permanent (static architecture). Noise in these signals will be called
gate control noise (GCN) in the following. It is to be noted that GCN
affects the most elementary gate operations, which work on the level
of physical qubits (using the term in the language of quantum error
correction). It is also to be noted that this type of decoherence
cannot be avoided through decoherence-free subspaces: Decoupling
logical qubits from GCN would also decouple them from the control
signals themselves. The decoherence due to GCN will be studied for
several exemplary architectures in the following.

There are many sets of fundamental gates allowing universal quantum
computation. For simplicity, the following will assume the controlled
phase gate as the fundamental two-qubit gate.  The controlled phase
gate is the basic two-qubit gate in Shor's quantum Fast Fourier
Transform algorithm \cite{shor94}. It is also the fundamental
two-qubit gate in one of the most elaborate proposals of a quantum
computing architecture so far \cite{taylo05}. A diagonal operator
similar to the controlled phase gate was recently implemented in a
demonstration gate consisting of two superconducting charge qubits
\cite{yamam03}. The fact that GCN in a controlled phase gate leads to
pure dephasing in the ordinary qubit basis $\{|0\rangle,\,
|1\rangle\}$ also helps to analyze the effects of decoherence separate
from other aspects of a quantum computer's dynamics.

{\em Fully switched array architecture.} --
The simplest architecture of a quantum computer consists of a fully
switched array, i. e., a circuit that contains a two-qubit quantum
gate for each pair of qubits. The number of control fields needed for
a fully switched array rises quadratically with the length $L$ of the
quantum register. The Hamiltonian term representing the array of
controlled-phase gates then is a double sum over all qubits,
\begin{equation}
\label{eq:pccoupling}
H_{\mathrm{fsa}} = \frac{1}{8} \sum_{jk} (\Phi_{jk} + \Xi_{jk})\, {\sf Z}_j {\sf Z}_k
,
\end{equation}
where ${\sf Z}_j$ denotes a diagonal Pauli matrix representing the $j$-th
qubit in the register. Here a distinction is made between nominal
control fields $\Phi_{jk}(t)$ and an additive, unbiased Gaussian noise
$\Xi_{jk}(t)$. The noise arises from thermal and quantum fluctuations
of the control circuit, which acts as a thermal reservoir. The
decoherence rate then has the usual dependence on the noise power
density $S(\omega)$ and the pointer variable,
\begin{equation}
\Gamma = \frac{1}{2\hbar^2} S(0) (Q-Q')^2
\label{eq:gamma}
,
\end{equation}
where $Q$ and $Q'$ denote eigenvalues of the pointer variable which
act as labels of off-diagonal elements of the density matrix.

For the case of {\em uniform} noise $\Xi(t)$ from a central source
(independent of $j$ and $k$) there is a single pointer variable $Q =
M^2/2$, where $M=\sum_j m_j$ is the $z$ component of the ``total
spin'' of the quantum register. Using an ohmic spectral density
\cite{weiss99} $J_{\mathrm{fsa}}(\omega) = s_\eta \omega
e^{-\omega/\omega_{\mathrm{c}}}$, which is characteristic of low-frequency
circuit noise, this leads to the decoherence rate
\begin{equation}
\label{eq:fsaurate}
\Gamma_{\mathrm{fsa,u}} = \frac{s_\eta k_{\mathrm{B}} T }{4\hbar^2}
 (M^2 - M'^2)^2
.
\end{equation}

The single most important feature of (\ref{eq:fsaurate}) is the quartic
dependence of $\Gamma_{\mathrm{fsa,u}}$ on $M$ and $M'$, even worse
than the quadratic dependence in ordinary superdecoherence. It is
easily verified that there is a large class of states for which the
decoherence rate grows much faster than linearly with the register
length $L$. This has serious implications on the arguments on which
proposals for error correction and fault-tolerant methods rely. In
particular, concatenation of error correction mechanisms
\cite{knill98} may turn out to be self-defeating after nesting very
few levels, since the resulting size of the quantum computer may drive
the decoherence rate, and therefore also the error rates above the
required thresholds.

GCN thus constitutes a universal decoherence mechanism which violates
the assumption of size-independent error rates, possibly invalidating
current strategies proposed against decoherence in quantum computers.

The situation looks somewhat better if the noise forces $\Xi_{jk}$ are
{\em statistically independent}. Decoherence is described by a set of
pointer variables $Q_{jk} = m_j m_k/2$. However, the decoherence rate
is again not bounded by any linear function of the register length,
\begin{equation}
\label{eq:fsairate}
\Gamma_{\mathrm{fsa,i}} = \frac{s_\eta k_{\mathrm{B}} T }{16\hbar^2} (L-N_{\mathrm{d}}) N_{\mathrm{d}}
,
\end{equation}
where $N_{\mathrm{d}} = \sum_j m_j - m_j'$ is the number of qubits for which
the left and right labels of the density matrix differ (Hamming
distance). For matrix elements with $N_{\mathrm{d}} \approx L/2$ the rate
$\Gamma_{\mathrm{fsa,i}}$ grows quadratically with the register length
$L$. Again the error rate per gate will grow with system size, with
detrimental consequences for quantum error correction. 

{\em Bus architecture.} --
The fully switched array not only has unfavorable decoherence
properties, it seems also very difficult to physically assemble
$L(L+1)/2$ hard-wired two-qubit gates for any sizeable $L$. A more
realistic circuit to perform gates with any qubit pair in a register
of Josephson qubits was suggested by Makhlin et al. \cite{makhl99}.
This proposal is similar to the bus architectures found in
conventional computers. One control signal per qubit is applied, the
nominal value of the signal being non-zero only for the single pair
currently performing a two-qubit gate. The Hamiltonian of the bus
coupling is of the general type
\begin{equation}
\label{eq:buscoupling}
H_{\mathrm{bus}} = \frac{1}{2} \left( \sum_{j=1}^L \frac{\varphi_j+\xi_j}{2} {\sf Z}_j \right)^2
,
\end{equation}
where again $\varphi_j$ is the nominal control signal and the unbiased
noise $\xi_j$ describes Gaussian fluctuations of the control
signal. When exactly two of the control fields are non-zero, a a
combination of the controlled-phase gate and single-qubit phase shifts
is performed; when all of the control fields vanish, all gates are
idle.  One salient feature of eq. (\ref{eq:buscoupling}) is the fact that
it contains a quadratic noise term, which is also present when all
gates are idle. This will lead to a shift in the inter-qubit couplings
even for unbiased noise.

The case of a single qubit with quadratic coupling to noise has
recently been studied by Makhlin and Shnirman \cite{makhl04}. Here we
explore lowest-order effects of GCN on an entire quantum register and
carefully explore their scaling with the register length $L$,
including the dynamic response of the reservoir linked to the noise
characteristics through the fluctuation-dissipation theorem. The
results presented in the following are valid in the limit of low
temperature.

\begin{figure}
\centerline{\includegraphics[width=\columnwidth]{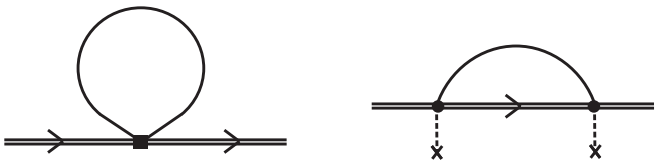}}
\caption[]{lowest-order diagrams for spurious coupling and decoherence
  for the bus-type inter-qubit coupling induced by gate control
  noise.}
\label{fig:diagrams}
\end{figure}

The two lowest-order irreducible diagrams are shown in
Fig. \ref{fig:diagrams}. A double line represents the propagation of
the quantum register, and a round (square) vertex corresponds to the
interaction terms $\frac{1}{4} \sum_{jk} \xi_j \xi_k {\sf Z}_j {\sf Z}_k$ and
$\frac{1}{4} \sum_{jk} \varphi_j \xi_k {\sf Z}_j {\sf Z}_k$ respectively. Curved
lines represent the correlation matrix $\langle \xi_j(t)
\xi_k(t')\rangle$.

The first diagram, fig. \ref{fig:diagrams}a, leads to no decoherence,
but a permanent {\em spurious coupling} $V_{\mathrm{sc}} = \frac{1}{16}
\sum_{j\neq k} \langle\xi_j \xi_k + \xi_k\xi_j\rangle {\sf Z}_j {\sf Z}_k$
between pairs of qubits. It depends on the characteristics
of the reservoir model. For concrete results, a specific model must be
assumed for the reservoir. As a minimal reservoir model a one-dimensional
field with linear dispersion, coupled locally to individual qubits,
has previously been employed \cite{unruh95}. To compare a different
case, a 3D reservoir field \cite{kless06} will also be considered, in
each case assuming the spectral density of the quantum noise
correlations at a single site to be of the ohmic type
$J_{\mathrm{bus}}(\omega) = \tau_\eta \omega \,\, e^{-\omega/\omega_{\mathrm{c}}}$.

The spectral density for different sites is
\begin{equation}
J_{jk}(\omega) = f(\omega r_{jk}/v) J_{\mathrm{bus}}(\omega)
\label{eq:jjk1d}
\end{equation}
with
\begin{equation}
f(x) = \left\{ 
\begin{array}{ll}
\cos(x) & ,\;\textrm{1D}\\
\sin(x)/x & ,\;\textrm{3D.}
\end{array}
\right.
\end{equation}
Here the prefactor $f$ reflects the time it takes a reservoir
excitation to travel between sites $j$ and $k$ with velocity $v$. Note
that in both cases $J_{\mathrm{bus}}(\omega)$ and $J_{jk}(\omega)$ are
identical in the low-frequency limit.

For the specific model chosen here, $V_{\mathrm{sc}}$ is of the form
$V_{\mathrm{sc}} = \frac{1}{8} \sum_{j\neq k} \mu^{(\mathrm{sc})}_{jk} {\sf Z}_j {\sf Z}_k$
with
\begin{equation}
\mu^{(\mathrm{sc})}_{jk} = \frac{\hbar}{\pi} \omega_{\mathrm{c}}^2 \tau_\eta
 \,g(\omega_{\mathrm{c}}r_{jk} / v)
\end{equation}
and
\begin{equation}
 g(x) = \left\{
\begin{array}{cl}
(1-x^2)/(1+x^2)^2 & ,\;\textrm{1D}\\
1/(1+x^2) & ,\;\textrm{3D.}
\end{array}
\right.
\end{equation}

This result shows that the spurious coupling $V_{\mathrm{sc}}$ is a necessary
feature of the bus architecture, whose precise strength and range
depend on details of the reservoir. It cannot be expected to vanish in
the case of a more realistic, structured environment. The spurious
coupling compromises the fidelity not only of individual gates, but of
the quantum computer as a whole. Correcting the spurious coupling,
however, seems feasible in principle since the coupling is not time
dependent.

The diagram shown in fig. \ref{fig:diagrams}b describes transient
phenomena induced by gate operation (two non-zero values $\varphi_j$).
Assuming gate operation to be slow compared to the reservoir
fluctuations, $\varphi_j$ can be taken constant when computing the
decoherence rate during gate operation as well as a transient
four-qubit coupling associated with the diagram shown in
Fig. \ref{fig:diagrams}b.

For decoherence, the reservoir properties enter only in the
low-frequency, long-wave limit, which simplifies further
considerations.  $J_{jk}(\omega)$ becomes independent of its indices
$j$, $k$ in the limit $\omega \to 0$, and no distinction needs to be
made between the 1D and 3D cases. A single pointer variable describes
decoherence during performance of a gate. It is the sum of two
non-zero terms representing the qubit pair of the active gate,
\begin{equation}
Q = M \sum_j\varphi_j m_j
\label{eq:point}
.
\end{equation}
In the ohmic case considered here, the decoherence rate is given by
\begin{equation}
\Gamma_{\mathrm{bus}} = \tau_\eta k_{\mathrm{B}} T (Q-Q')^2 / \hbar^2
,
\label{eq:gammaohm}
\end{equation}
which scales quadratically with $L$. Again, eqs. (\ref{eq:point}) and
(\ref{eq:gammaohm}) indicate superdecoherence.

The diagram of fig. 1b also introduces a transient shift of the energy
levels of the quantum registers, which is required by the
fluctuation-dissipation theorem. It is the sum of many inter-qubit
coupling terms,
\begin{equation}
\Delta E_{\mathrm{tr}}
 = \frac{1}{2} \sum_{jkln} \varphi_j \varphi_l \mu^{(\mathrm{tr})}_{kn} m_j m_k
m_l m_n ,
\label{eq:etr}
\end{equation}
where
\begin{eqnarray}
\mu^{(\mathrm{tr})}_{kn} = \mu^{(\mathrm{tr})}_{nk} &=&
- \frac{1}{\hbar} \mathop{\mathrm{Im}} \int\limits_0^\infty d \tau\,
 \langle \xi_k(\tau)\xi_n(0) - \xi_n(0)\xi_k(\tau) \rangle \nonumber\\
&=&
\frac{2}{\pi} \int\limits_0^\infty
 \frac{d \omega}{\omega} J_{kn}(\omega)
.
\end{eqnarray}

In the case $k=n$, the corresponding term in Eq. (\ref{eq:etr}) affects
only the pair of qubits involved in the active gate. The energy shift
enhances the effect of the gate control fields $\varphi_j$ and
$\varphi_l$ by a factor $1 + L\tau_\eta \omega_{\mathrm{c}} / 4\pi $. This
can be compensated by re-calibrating the gate control fields. For
$k\neq n$ gate operation introduces a transient four-qubit coupling
which may be difficult to compensate. The qubits labeled $j$ and $l$
form the active gate while $k$ and $n$ are arbitrary qubits which are
also involved in the four-qubit interaction. The strength of this
interaction depends only on the separation $r_{kn}$ of the two qubits
$k$ and $n$.  Its details are determined by the characteristics of the
reservoir. For the 1D and 3D models, one finds
\begin{equation}
\mu^{(\mathrm{tr})}_{kn} = 2\tau_\eta \omega_{\mathrm{c}} \,h(\omega_{\mathrm{c}} r_{kn}/v)
/ \pi
\end{equation}
with
\begin{equation}
h(x) = \left\{
\begin{array}{ll}
1/(1+x^2) & ,\;\textrm{1D} \\
\arctan(x)/x & ,\;\textrm{3D}
\end{array}
\right.
.
\end{equation}
It is therefore impossible to perform a two-qubit gate without
disturbing other qubits. In particular, the 3D case leads to an
interaction which decays only as $1/r_{kn}$ with distance.

{\em Hypercube and processor core architectures.} --
A variant of the fully switched architecture is the hypercube
architecture. Here the graph describing connected qubit pairs is given
by the edges of a $\log_2 L$-dimensional hypercube. Connected qubits
are assumed to be capable of swap operations and one of the usual
two-qubit gates as well as single-qubit gates. With these operations,
this architecture is clearly capable of universal quantum computation.
The decoherence mechanism is the same as in the fully switched
architecture, but the number of qubit pairs to be summed over is much
smaller.  In the case of independent noise sources, the decoherence
rate is proportional to the total number of $\frac{1}{2} L \log_2 L$
physical gates, which may be acceptable for quantum error
correction. A central source of GCN, however, would lead to much
faster, probably uncorrectable decoherence in this model.

The processor core model of Yung et al. \cite{yung06} leads to a
similar picture. In the processor core, only `always on' couplings are
used for $L$ qubits. Since these couplings need not rely on signals
from an external circuit, they are immune to GCN. Each of the qubits
in the processor core can be swapped with one qubit in a storage bank,
whose elements do not interact directly. This leads to a universal
architecture with only $L$ gates vulnerable to GCN. Accordingly, the
decoherence rate in the processor core model is proportional to $L$
($L^2$) for independent noise sources (a global noise source).

{\em Conclusions.} --
Noise in gate control fields can profoundly alter the characteristics
of a quantum computer in a way not described by the failure of
individual two-qubit quantum gates. It causes a new decoherence
mechanism whose rate often scales quadratically with system size or
worse, leading to superdecoherence. This is due to the fact that it
introduces pointer variables whose upper bound grows faster than
linearly as a function of the register length $L$.  The finding of
superdecoherence raises the serious question whether known error
correction methods are scalable in the presence of gate control
noise. The GCN decoherence problem cannot be solved by resorting to
logical qubits in decoherence-free subspaces: States which are
decoupled from the noise forces are also decoupled from the control
fields. The fully switched and bus architectures appear particularly
vulnerable to gate control noise, while the hypercube architecture and
the processor core architecture seem less affected. A central noise
source, however, will lead to superdecoherence in all of the
architectures considered.

The preceding analysis is appropriate for the static architectures
typical of solid-state quantum computers. In this setting, GCN equally
can affect `idle' gates as well as `active' gates. Architectures
employing moving qubits may be successful in isolating qubits from
gate control noise when they are idle. They are not described by any
of the cases discussed here and may therefore be immune to the
decoherence mechanisms described. The investigation of gate control
noise for these quantum computing architectures should be of great
interest.

\acknowledgments This research was supported by Deutsche
Forschungsgemeinschaft (DFG) under Sonderforschungsbereich 382.

\bibliography{gcolldec,Juergen}

\begin{thebibliography}{10}

\bibitem{feynm85}
R.~P. Feynman, Opt. News {\bf 11},  11  (1985).

\bibitem{unruh95}
W.~G. Unruh, Phys. Rev. A {\bf 51},  992  (1995).

\bibitem{zurek91}
W.~H. Zurek, Physics Today {\bf 44},  36  (1991).

\bibitem{stean96}
A.~M. Steane, Phys. Rev. Lett. {\bf 77},  793  (1996).

\bibitem{knill97a}
E. Knill and R. Laflamme, Phys. Rev. A {\bf 55},  900  (1997).

\bibitem{knill00}
E. Knill, R. Laflamme, and L. Viola, Phys. Rev. Lett. {\bf 84},  2525  (2000).

\bibitem{shor96}
P.~W. Shor,  in {\em Proceedings of the 37th Annual Symposium on the
  Foundations of Computer Science}, IEEE Computer Society (IEEE Comp. Soc.
  Press, Los Alamitos, CA, 1996), pp.\ 56--65.

\bibitem{knill98}
E. Knill, R. Laflamme, and W.~H. Zurek, Science {\bf 279},  342  (1998).

\bibitem{knill05}
E. Knill, Nature {\bf 434},  39  (2005).

\bibitem{svore05}
K.~M. Svore, B.~M. Terhal, and D.~P. DiVincenzo, Phys. Rev. A {\bf 72},  022317
   (2005).

\bibitem{aharo06}
D. Aharonov, A. Kitaev, and J. Preskill, Phys. Rev. Lett. {\bf 96},  050504
  (2006).

\bibitem{kless06}
R. Klesse and S. Frank, Phys. Rev. Lett. {\bf 95},  230503  (2006).

\bibitem{novai06}
E. Novais and H.~U. Baranger, Phys. Rev. Lett. {\bf 97},  040501  (2006).

\bibitem{zanar97}
P. Zanardi and M. Rasetti, Phys. Rev. Lett. {\bf 79},  3306  (1997).

\bibitem{lidar98}
D.~A. Lidar, I.~L. Chuang, and K.~B. Whaley, Phys. Rev. Lett. {\bf 81},  2594
  (1998).

\bibitem{shor94}
P.~W. Shor,  in {\em Proceedings of the 35th Annual Symposium on the
  Foundations of Computer Science}, IEEE Computer Society, edited by S.
  Goldwasser (IEEE Comp. Soc. Press, Los Alamitos, CA, 1994), pp.\ 124--134.

\bibitem{taylo05}
J.~M. Taylor {\it et~al.}, Nat. Phys. {\bf 1},  177  (2005).

\bibitem{yamam03}
T. Yamamoto, Y.~A. Pashkin, O. Astafiev, and Y. Nakamura, Nature {\bf 425},
  941  (2003).

\bibitem{weiss99}
U. Weiss, {\em Quantum Dissipative Systems}, 2nd ed. (World Scientific,
  Singapore, 1999).

\bibitem{makhl99}
Y. Makhlin, G. Sch{\"o}n, and A. Shnirman, Nature {\bf 398},  305  (1999).

\bibitem{makhl04}
Y. Makhlin and A. Shnirman, Phys. Rev. Lett. {\bf 92},  178301  (2004).

\bibitem{yung06}
M.-H. Yung, S.~C. Benjamin, and S. Bose, Phys. Rev. Lett. {\bf 96},  220501
  (2006).

\end{thebibliography}

\end{document}